\documentclass{article}

\usepackage{amsmath}
\usepackage{amssymb}
\usepackage{amsfonts}
\usepackage{epsfig}

\newcounter{NN}
\newtheorem{proposition}[NN]{Proposition}
\newtheorem{theorem}[NN]{Theorem}
\newtheorem{algorithm}[NN]{Algorithm}
\newtheorem{definition}[NN]{Definition}
\newtheorem{lemma}[NN]{Lemma}
\newtheorem{corollary}[NN]{Corollary}

\def\proof{\noindent {\bf Proof:} }
\def\qed{{\hfill $\square$\\ \noindent}}

\def\N{\mathbb{N}}

\def\AR{{\mathcal A}}
\def\BR{{\mathcal B}}

\def\SS{{\mathcal S}}

\def\pr{{\text{pr}}}
\def\cpr{{\mathfrak{pr}}}
\def\D{{\mathfrak{D}}}

\def\w#1{{\widehat{#1}}}

\begin{document}
\bibliographystyle{plain}
\title{Symmetry condition in terms of Lie brackets.}
\author{Peter H.~van der Kamp}
\date{Department of Mathematics and Statistics, \\
La Trobe University, Victoria, 3086, Australia\\[3mm]
Email:\ peterhvanderkamp@gmail.com }

\pagestyle{plain} \maketitle

\begin{abstract}
A passive orthonomic system of PDEs defines a submanifold in the
corresponding jet manifold, coordinated by so called parametric
derivatives. We restrict the total differential operators and the
prolongation of an evolutionary vector field $v$ to this
submanifold. We show that the vanishing of their commutators is
equivalent to $v$ being a generalized symmetry of the system.
\end{abstract}

\section{Concerning the status of this preprint}
After writing this preprint, I learnt (from an anonymous referee)
that the Lie-bracket criterion is valid for all systems, not only
orthonomic ones. Vinogradov-style derivation can be found for instance
in \cite[Chapter 4, $\S 3$]{KV}. The `if' part of the criterion follows
from $\S 3.3$, while lemma 3.6 in $\S 3.4$ gives the `only if' part.

Therefore, the present result is not the most general one. Still the
reader may appreciate its Van der Kamp-style derivation. The intrinsic
differential operators we will define, can be useful in practical
situations.

\section{The standard symmetry condition}
In the majority of cases where exact solutions of differential
equations can be found, the underlying property is a (continuous)
symmetry of the equation \cite{ste,olv}. And, in the theory of integrable
equations, the recognition and classification methods based on the
existence of symmetries have been particular successful
\cite{MSS, MR99g:35058, TsWo, vdK, MNW}.

A symmetry-group transforms one solution of an equation to another
solution of the same equation. Although this idea goes back to Sophus Lie,
we refer to \cite{olv} for a
good introduction to the subject, numerous examples, applications and
references.  And we quote: 'The great power of Lie group theory lies
in the crucial observation that one can replace the complicated,
nonlinear conditions for the invariance of the solution set of an
equation under the group transformations by an equivalent linear
condition of infinitesimal invariance under the corresponding
infinitesimal generators of the group action' \cite{olv}. In this paper
we provide a characterization of symmetries that is different from the
standard one, generalizing a similar characterization in the special
setting of ordinary differential equations \cite[eq. (3.35)]{ste},
and evolution equations \cite[Prop. 5.19]{olv} to the setting of
passive orthonomic systems.

The natural framework in which symmetries of differential equations are
studied is the so called jet-manifold $M$. Coordinates on $M$ consist
of $p$ independent variables $x_i$, $q$ dependent variables
$u^\alpha$ and the derivatives of the dependent variables, which are
denoted using multi-index notation, e.g.
\[
u^2_{1,0,3}=\frac{\partial^4 w}{\partial r \partial t^3}
\]
when $x=(r,s,t)$ and $u=(v,w)$. A typical point $z\in M$ is
$z=(x_i,u^\alpha,u^\alpha_K)$. The ring of smooth functions on $M$
will be denoted $\AR$. To indicate functional dependence of $f\in
\AR$ we simply write $f(z)$. Thus the system $\Delta(z)=0,\ \Delta\in \AR^n$
is a system of $n$ PDEs.

The action of a Lie group is defined on the space of dependent and
independent variables, and then prolonged to an action on the
jet manifold. Likewise the infinitesimal generator of the
symmetry group is obtained by prolongation from an infinitesimal
vector field on the base manifold. It turns out that any symmetry has
an evolutionary representative \cite[Prop. 5.5]{olv}. In
terms of the total differential operators
\begin{equation} \label{td}
D_i=\frac{\partial}{\partial x_i} + \sum_{\alpha,K} u_{Ki}^\alpha
\frac{\partial}{\partial u_K^\alpha},\ i=1,\ldots,p,
\end{equation}
the prolongation $\pr_Q$ of an evolutionary vector field
$\nu_Q=\sum_\alpha Q^\alpha
\partial/\partial u^\alpha$ is
\begin{equation} \label{pr}
\pr_Q=\sum_{\alpha,K} D_K Q^\alpha
\frac{\partial}{\partial u_K^\alpha}.
\end{equation}
A simple computation shows that these derivations on $\AR$
commute among each other, we have $[D_i,D_j]=0$, $i,j=1,\ldots,p$ and
\begin{equation} \label{proper}
[D_i,\pr_Q]=0,\quad i=1,\ldots,p,\ Q\in \AR.
\end{equation}
In fact, up to a linear combination of translational fields $\partial/\partial x_i$,
evolutionary vector fields are uniquely determined by property (\ref{proper}),
cf. \cite[Lemma 5.12]{olv}.

The condition of infinitesimal invariance, {\em the standard symmetry
condition}, is \cite[Theorem 2.31]{olv}
\begin{equation} \label{infinv}
\pr_Q\Delta\equiv 0 \text{ mod } \Delta
\end{equation}
in which case $\nu_Q$, or the tuple $Q\in\AR^n$ itself, is called a
(generalized) symmetry of the system $\Delta=0$. The tuple $Q$ is a trivial
symmetry if $Q\equiv 0 \text{ mod } \Delta$, which defines an
equivalence relation on the space of symmetries. In section \ref{pos} we
show this is well defined for {\em passive orthonomic systems}.
We restrict $Q$ to be a function $Q\in\BR$ on the sub-manifold of the
jet-manifold defined by our system of PDEs. Although this is a more than
standard procedure, restricting the derivations to act on this
sub-manifold is not standard at all, except possibly in the settings
of ODEs and evolution equations. This is, at least from a philosophical
point of view, not fully satisfying.

In section \ref{isc}, for any passive orthonomic system of partial
differential equations, we define intrinsic total differential
operators $\D_i$ and an intrinsic prolongation $\cpr_Q$,
which are derivations on the sub-space $\BR$. Subsequently we
show that the vanishing of the Lie brackets
\[
[\D_i,\cpr_Q]=0,\quad i=1,\ldots,p,\ Q\in\BR
\]
is equivalent to $Q$ being a symmetry.

\section{Passive orthonomic systems} \label{pos}
Restricting to the sub-manifold is particularly easy when dealing
with orthonomic systems, in which case this amounts to using the
equations as substitution rules. However, in general the order of
substituting and differentiating does matter, one encounters
integrability conditions. For example, for the system $u_x=X, u_y=Y$
to be formally integrable we need $D_yX=D_xY$. In general, a finite
number of integrability conditions suffices to make the system
formally integrable, in which case the system is called passive.

The idea of a passive orthonomic system is the main idea behind Riquier's
existence theorems \cite{tho} and the corresponding algorithms
for solving systems of PDEs due to Janet \cite{jan}. Riquier-Janet theory
extended the works of Cauchy and Kovalevskaya, it takes a prominent place
in computer algebra applied to PDE theory \cite{rwb}, and it has lead to
important developments in polynomial elimination theory \cite{gebl}. The passive
orthonomic system was the predecessor of what is now called an involutive
system of PDEs. For our purpose, the concept of involutivity does not play
a role. We adopt a similar philosophy as in \cite{marv}, and stick to the
setting of passive orthonomic systems. In that paper an efficient algorithm is
given by which any orthonomic system can be made passive by construction
of a sufficient set integrability conditions free of redundancies \cite{marv}.

Let $\N=\{0,1,2,\ldots\}$ and $\N_q=\{1,2,\ldots,q\}$. We denote the
$i$-th component of $J\in\N^p$ by $J_i$ and addition in $\N^p$ is
denoted by concatenation. A set of basis vectors for $\N^p$ is given
by $\{1,2,\ldots,p\}$, where $i_j=1$ when $i=j$ and $i_j=0$
otherwise. Thus, with $J,K\in \N^p$, we have $(JK)_i=J_i+K_i$, and in
particular $(Kj)_i$ equals $K_i$ or, when $j=i$, $K_i+1$. Also, when $L=JK$
we write $J=L/K$. Since total differential operators commute we can define a
multi-differential operator $D_K$ as
\begin{equation} \label{DK}
D_K = D_1^{K_1} D_2^{K_2} \cdots D_p^{K_p},
\end{equation}
and we have $D_K u^j=u^j_K$.

Choose $n$ points $(i^\alpha,J^\alpha) \in \N_q\times\N^p$, with nonzero
$J^\alpha$, $\alpha=1,\ldots,n$. A derivative $u^j_K$ is called {\em principal}
if there exist $L\in \N^p$ such that $(j,K)=(i^\alpha,J^\alpha L)$
for some $\alpha$. The remaining ones are called {\em parametric}.
The set of all $(j,K)$ such that $u^j_K$ is parametric is denoted
$\SS$, and the subspace of $\AR$ consisting of smooth functions of the
parametric derivatives is denoted $\BR$.

We also choose a ranking $\leq$ on $\N_q\times\N^p$, that is, a total order
relation which is positive:
\[
\forall L, (j,K) \leq (j,KL),
\]
and, compatible with differentiation:
\[
(i,J)\leq (j,K) \Leftrightarrow (i,JL)\leq (j,KL),
\]
cf. \cite{ogol,rust}.

We consider systems of $n$ partial differential equations, with $\alpha=1,\ldots,n$,
\begin{equation} \label{hde}
u^{i^\alpha}_{J^\alpha} = P^\alpha, \qquad P^\alpha\in \BR.
\end{equation}
The system (\ref{hde}) will be written shortly $\Delta=0$, where $\Delta^\alpha = u^{i^\alpha}_{J^\alpha} - P^\alpha$.
We make the following assumptions:
\begin{enumerate}
\item[$i)$] the $P^\alpha$ only depend on $u^j_K$ with $(j,K)<(i^\alpha,J^\alpha)$, and
\item[$ii)$] $(i^\alpha,J^\alpha K)=(i^\beta,J^\beta L) \Rightarrow D_KP^\alpha=D_LP^\beta$.
\end{enumerate}
Such systems are called {\em passive orthonomic systems}. Their crucial property is
that for any $Q\in\AR$, there is a unique $\w{Q}\in\BR$ such that $\w{Q}\equiv Q \mod \Delta$.
This $\w{Q}$ can be obtained from the following reduction algorithm.
\begin{algorithm} \label{alg} {\em Input:} Expression $Q\in\AR$. {\em Output:} Expression $\w{Q}\in\BR$.
\begin{itemize}
\item[$\star$] if no principal derivative appears in $Q$ then return $Q$
\item[$\star$] let $u^j_K$ be the $\leq$-highest principal derivative appearing in $Q$, and let
$\alpha,L$ be such that $j,K=i^\alpha,J^\alpha L$
\item[$\star$] substitute $u^j_K=D_LP^\alpha$ in $Q$ and call the result $R$
\item[$\star$] return $\w{R}$
\end{itemize}
\end{algorithm}
The algorithm terminates because the highest principal derivative of $R$, if
it exists, is $\leq$-smaller than $u^j_K$, due to assumption i). And,
a different choice of $\alpha$ wouldn't change the result because of
assumption ii).

The following lemma states that differentiation is compatible with the above reduction
$\AR\rightarrow\BR$, cf. \cite[Theorem 4.8]{marv}.
\begin{lemma} \label{dcr}
For any $Q\in\AR$ we have
\[
\w{D_KQ}=\w{D_K\w{Q}}
\]
\end{lemma}
\proof Using a modified version of Algorithm \ref{alg} we can write $Q=\w{Q}+R(\Delta)$,
where $R$ is some differential function of $\Delta$ such that $R(0)=0$.
Clearly $\w{D_K R(\Delta)}$ vanishes. \qed

\section{The intrinsic symmetry condition} \label{isc}
\begin{definition} \label{wD}
We define an {\em intrinsic multi-differential operator} $\D_K: \BR \rightarrow \BR$ by
\[
\D_K P = \w{D_K P}, \qquad P \in \BR
\]
\end{definition}
From this definition and Lemma \ref{dcr} we obtain the following properties.
\begin{proposition} \label{cwc}
Intrinsic multi-differential operators are compatible with concatenation, $\D_K\D_L=\D_{KL}$.
\end{proposition}
\proof We have $\D_K\D_L P = \w{D_K\w{D_L P}} = \w{D_KD_L P} = \w{D_{KL} P} = \D_{KL} P.$ \qed
\begin{corollary} \label{ADK}
We have the analogue of equation (\ref{DK}), $\D_K = \D_1^{K_1} \D_2^{K_2} \cdots \D_p^{K_p}$.
\end{corollary}
\begin{corollary}
Intrinsic total differential operators commute, $[\D_i,\D_j]=0$.
\end{corollary}
We would like to have a more intrinsic characterization of $\D_i$,
that is, without reference to any principal derivative or total differential
operator. For $L\in\N^p$ we denote $\SS_L=\{(\alpha,K): (\alpha,KL) \in\SS\}$, which is a subset
of $\SS$. From equation (\ref{td}) and Definition \ref{wD} it follows that
\begin{equation} \label{itdo}
\D_j = \frac{\partial }{\partial x_j} + \sum_{(k,L)\in\SS_j} u^k_{Lj} \frac{\partial }{\partial u^k_L}
          + \sum_{(i^\alpha,J^\alpha M/j) \in \SS\setminus \SS_j} \D_{M} P^\alpha \frac{\partial }{\partial u^{i^\alpha}_{J^\alpha M/j}}.
\end{equation}
We note that when $(k,L)\in \SS\setminus \SS_j$ there exist $\alpha\in\N_n$, $M\in\N^p$ such that $(k,Lj)=(i^\alpha, J^\alpha M)$,
and, for any $\beta\in\N_n$, $N\in\N^p$ such that $(k,Lj)=(i^\beta, J^\beta N)$
we have $i^\alpha=i^\beta$ and $\D_{M} P^\alpha=\D_{N} P^\beta$. Due to Corollary \ref{ADK} equation (\ref{itdo})
provides a recursive schema for intrinsic total differentiation.
\begin{proposition}
The recursive schema (\ref{itdo}) for $\D_i$ is well defined.
\end{proposition}
\proof
We show the schema terminates using transfinite induction. For any $Q\in\AR$, to evaluate $\D_j Q$, apart from
some differentiation and multiplications, we need to evaluate a finite number of expressions $\D_I P^\alpha$.
We assume that $\D_L P^\beta$ can be evaluated for all $(i^\beta,J^\beta L)<(i^\alpha,J^\alpha I)$.
Suppose $P^\alpha$ depends on $u^j_K$. That implies $(j,K)<(i^\alpha,J^\alpha)$. Suppose there are $\beta\in\N_n$ and
$L\in\N^p$ such that $(j,KI)=(i^\beta,J^\beta L)$. Then, to evaluate $\D_I P^\alpha$ one may need to evaluate
$\D_L P^\beta$. By the induction hypothesis this can be done.
\qed

\begin{definition} \label{ip}
We define {\it intrinsic prolongation}, denoted $\cpr_Q:\BR\rightarrow\BR$, of an evolutionary vector field
$\nu_Q$ with $Q\in\BR$ by
\[
\cpr_Q P = \w{\pr_Q P}, \qquad P \in \BR.
\]
\end{definition}
From equation (\ref{pr}) and Definition \ref{ip} we get the intrinsic formula
\[
\cpr_Q = \sum_{j,K\in\SS} \D_K Q^j \frac{\partial}{\partial u^j_K}.
\]

We now state and prove our main theorem.
\begin{theorem}
A tuple $Q\in\BR $ is a symmetry of equation (\ref{hde}) iff
\[
[\D_j,\cpr_Q]=0
\]
for all $j$.
\end{theorem}
\proof
\begin{itemize}
\item[$\Leftarrow$]
We calculate $\pr_Q \Delta^\alpha$ modulo $\Delta$
\begin{eqnarray}
\w{\pr_Q \Delta^\alpha} &=& \w{D_{J^\alpha}Q^{i^\alpha}}-\w{\pr_Q P^\alpha} \nonumber \\
&=& \D_{J^\alpha}Q^{i^\alpha} -\cpr_Q P^\alpha \label{rhs}
\end{eqnarray}
Next we calculate the commutator $[\D_j,\cpr_Q]$.
Neglecting second order derivatives, we get
\[
\D_j \cpr_Q = \sum_{(k,L)\in\SS} \D_{Lj} Q^k \frac{\partial}{\partial u^k_L},
\]
and
\[
\cpr_Q \D_j = \sum_{(k,L)\in\SS_j} \D_{Lj} Q^k \frac{\partial}{\partial u^k_L}
+\sum_{(i^\alpha,J^\alpha M/j)\in\SS\setminus\SS_j} \cpr_Q \D_M P^\alpha \frac{\partial}{\partial u^{i^\alpha}_{J^\alpha M/j}}.
\]
Hence we get
\[
[\D_j,\cpr_Q]=\sum_{(i^\alpha,{J^\alpha M/j})\in\SS\setminus \SS_j}
\left( \D_{J^\alpha M}Q^{i^\alpha}-\cpr_Q \D_M P^\alpha \right)
\frac{\partial}{\partial u^{i^\alpha}_{J^\alpha M/j}}.
\]
Suppose that $J^\alpha_j\neq 0$. Then the action of the above
vector field on $u^{i^\alpha}_{J^\alpha/j}$ yields the right hand side of
equation (\ref{rhs}). Since we have chosen $J^\alpha\neq 0\in\N^p$ this proves our case.
\item[$\Rightarrow$]
Suppose $\D_{J^\alpha}Q^{i^\alpha} = \cpr_Q P^\alpha$. Then
\[
[\D_j,\cpr_Q] =
\sum_{(i^\alpha, J^\alpha M/j )\in\SS\setminus \SS_j}
\left[ \D_M , \cpr_Q \right] P^\alpha
\frac{\partial}{\partial u^{i^\alpha}_{J^\alpha M/j}}.
\]
We will prove that $\left[ \D_M , \cpr_Q \right] P^\alpha =0$ for all $\alpha\in\N_n$ and $M\in\N^p$.
The statement is certainly true for $M=0$. Assume that $\left[ \D_N , \cpr_Q \right] P^\beta =0$ for all
$(i^\beta,J^\beta N) < (i^\alpha,J^\alpha M)$. When $M\neq 0$ there exists $j$ such that $M/j\in\N^p$. We write
\[
\left[ \D_M , \cpr_Q \right] P^\alpha = \D_j \left[ \D_{M/j} , \cpr_Q \right] P^\alpha + \left[ \D_j , \cpr_Q \right] \D_{M/j} P^\alpha
\]
The first term is zero by the induction hypothesis, so we concentrate on the second, which is
\[
\sum_{(i^\beta,J^\beta N)\in S\setminus S_j} \left[ \D_N , \cpr_Q \right] P^\beta \frac{\partial}{\partial u^{i^\beta}_{J^\beta N/j} }
\D_{M/j} P^\alpha.
\]
Suppose $P^\alpha$ depends on $u^k_L$. Then $(k,L)<(i^\alpha,J^\alpha)$. The function $\D_{M/j} P^\alpha$ may depend on
the derivative $u^k_{LM/j}$ (namely, if $u^k_{LM/j}\in\SS$, otherwise it depends on smaller derivatives). But when
$(i^\beta,J^\beta N/j)\leq (k,LM/j)<(i^\alpha,J^\alpha M/j)$ by the induction hypothesis $\left[ \D_N , \cpr_Q \right] P^\beta$
vanishes. \qed

\end{itemize}

\section{Discussion}
We have show that the infinitesimal condition $[\D_j,\cpr_Q]=0$, for all $j$, is completely
equivalent to the more standard infinitesimal symmetry condition $\pr_Q\Delta\equiv 0$ mod $\Delta$
in the setting of passive orthonomic systems (\ref{hde}). These conditions are sufficient
for the corresponding group of transformations to be a symmetry, in the sense of transforming solutions
into solutions \cite[Theorem 2.31]{olv}. In general, one needs to make non-degeneracy assumptions on
the systems $\Delta$ to ensure that they are also necessary, see \cite[Section 2.6 and 5.1]{olv}. However,
in the setting of passive orthonomic systems local solvability is ensured by the existence theorems of
Riquier and Janet \cite{tho,jan}. This means that infinitesimal methods can be used for classifying
all the symmetries of passive orthonomic systems.

We do not propose that the bracket condition should replace the prolongation condition for practical
calculations. On the other hand, the intrinsic differential operators do provide a convenient way of
restricting expressions to the sub-manifold defined by the equation.
For example, consider the Boussinesq equation
\[
u_{tt}=P,\ P=u_{xxxx}-2u_xu_{xx}
\]
If we let the prolongation $\pr_Q$, with $Q=u_{txx}-u_tu_x$, act on it we find
\begin{eqnarray*}
\pr_Q(u_{tt}-P)&=&u_{tttxx}-2u_{t}u_{xx}^2-6u_{tx}u_{x}u_{xx}+6u_{txxx}u_{xx}-2u_{txx}u_{x}^2\\
                 &&-2u_{t}u_{x}u_{xxx}+3u_{txxxx}u_{x}+6u_{txx}u_{xxx}+4u_{tx}u_{xxxx}\\
                 &&+u_{t}u_{xxxxx}-u_{txxxxxx}-2u_{tx}u_{tt}-u_{ttx}u_{t}-u_{ttt}u_{x} .
\end{eqnarray*}
To see whether this vanishes modulo the equation we have to substitute iteratively $u_{tt}=P$ whenever
we see two $t$-derivatives. An alternative is to work with intrinsic differential operators. The parametric
derivatives are $u,u_x,u_{tx},u_{xx},\ldots,$ $u_{x^n},u_{tx^n},\ldots$. Clearly we have $\D_x=D_x$. For the intrinsic total $t$-derivative we have
\[
\D_t=\frac{\partial}{\partial t} + \sum_n u_{tx^n} \frac{\partial}{\partial u_{x^n}} + \D_x^n P \frac{\partial}{\partial u_{tx^n}}
\]
which can straightforwardly be implement in any symbolic manipulation package. Without having to
perform any further simplifications it follows that
\[
\D_t^2 Q - \cpr_{Q}P = 0.
\]
We do not want to check that $[\D_t,\cpr_Q]=0$ using a computer since it involves infinite
vector-fields, which we can only deal with if we let them act on finite expressions.

The infinitesimal symmetry $Q$ is the first of an infinite hierarchy. The next member of
this hierarchy is given by
\[
R=u_{t}^2-\frac{8}{5}u_{xxxxx}+3u_{xx}^2+4u_{xxx}u_{x}-\frac{2}{3}u_{x}^3,
\]
In fact, the symmetries of this hierarchy commute modulo the Boussinesq equation, we have
$[\cpr_Q,\cpr_R]=0$. Using our bracket formulation of infinitesimal symmetry together with the
Jacobi-identity one can conclude that the bracket of two symmetries should again be a symmetry.
However, it is not a priori clear that this bracket is the intrinsic prolongation of something.

We have observed that up to some scaling, the 'spacial part' of $R$ equals the
right hand side of the integrable potential Kaup-Kupershmidt equation \cite[Equation 4.2.5]{MSS}
\[
u_t=u_{xxxxx}+10u_xu_{xxx}+\frac{15}{2} u_{xx}^2+\frac{20}{3} u_x^3,
\]
and the spatial parts of the other symmetries in the hierarchy of the Boussinesq equation are the
symmetries of the potential Kaup-Kupershmidt equation. At orders where the Kaup-Kupershmidt equation
has gaps in its hierarchy, the symmetries of Boussinesq do not have a spacial part.

Since the Fr\'{e}chet derivative of the Boussinesq equation is self-adjoint, its symmetries are also co-symmetries of the equation. Therefore they are in some sense dual to the Boussinesq equation.
In the discrete setting this kind of duality was first introduced in \cite{QCR}. Considering the dual
equation $u_{txx}=u_tu_x$, amongst its symmetries we find the hierarchy of the Sawada-Kotera equation
\cite{SK}
\[
u_t=u_{xxxxx}-5u_{xxx}u_x+\frac{5}{3}u_x^3.
\]
We note that a similar observation, using Lax-pairs, was made in \cite{HW}. These examples show the importance
of extending the techniques developed for establishing the integrability of evolution equations and classifying them, see \cite{MSS, MR99g:35058, TsWo, vdK, MNW} and references in there, to the realm of passive orthonomic systems, or beyond.

\subsection*{Acknowledgment} I sincerely thank Peter Olver for pointing me
in the direction of passive orthonomic systems. This research has been funded by the Australian Research Council through the Centre of Excellence for Mathematics and Statistics of Complex Systems. Also thanks
to the anonymous referee for bringing to my attention a more general result, see section 1.

\end{document}